# MANAGING AND ANALYSING SOFTWARE PRODUCT LINE REQUIREMENTS

Shamim Ripon[1], Sk. Jahir Hossain[1] and Touhid Bhuiyan[2]

[1]Department of Computer Science and Engineering, East West University, Bangladesh,
[2]Department of Software Engineering, Daffodil International University, Bangladesh

## ABSTRACT

*Modelling software product line (SPL) features plays a crucial role to a successful development of SPL. Feature diagram is one of the widely used notations to model SPL variants. However, there is a lack of precisely defined formal notations for representing and verifying such models. This paper presents an approach that we adopt to model SPL variants by using UML and subsequently verify them by using first-order logic. UML provides an overall modelling view of the system. First-order logic provides a precise and rigorous interpretation of the feature diagrams. We model variants and their dependencies by using propositional connectives and build logical expressions. These expressions are then validated by the Alloy verification tool. The analysis and verification process is illustrated by using Computer Aided Dispatch (CAD) system.*

## 1. INTRODUCTION

Designing, developing and maintaining a good software system is a challengingtask still in this 21st century. The approach of reusing existing good solutions fordeveloping any new application is now one of the central focuses of software engineers. Building software systems from previously developed components savescost and time of redundant work, and improves the system and its maintainability. A new software development paradigm, software product line [2], is emergingto produce multiple systems by reusing the common assets across the systems inthe product line. However, the idea of product line is not new. In 1976 Parnas [3]proposed modularization criteria and information hiding for handling productline.

Core assets are the basis for software product line. The core assets ofteninclude the architecture, reusable software components, domain models, requirements statements, documentation and specifications, performance model, etc. Different product line members may differ in functional and non-functional requirements, design decisions, run-time architecture and interoperability (component structure, component invocation, synchronization, and data communication), platform, etc. The product line approach integrates two basic processes:the abstraction of the commonalities and variabilities of the products considered (development for reuse) and the derivation of product variants from theseabstractions (development with reuse) [4].

The objective of this work is to provide an approach to modelling variants inthe domain model of a product line. In our approach, we initially consider a domain model which includes default domain view, a variant model and customization requirements. Default domain views describe typical system in a domain. Default domain views are the starting point for understanding the scope of the product line, i.e., the range of systems in the domain we wish to consider. We draw a model to represent the variants of a product line. The model contains all the variant related information required for customizing any product. After getting the requirements for any particular product of the product line, the product line model collects proper variant information from the variant model. A flexible variant configuration tool (FVC) interprets the variant model and customizes the default domain model by adapting and customizing the default domain according to the particular product requirements. Fig. 1 gives a top level view of the targeted variant model along with its position and activity with product line model.

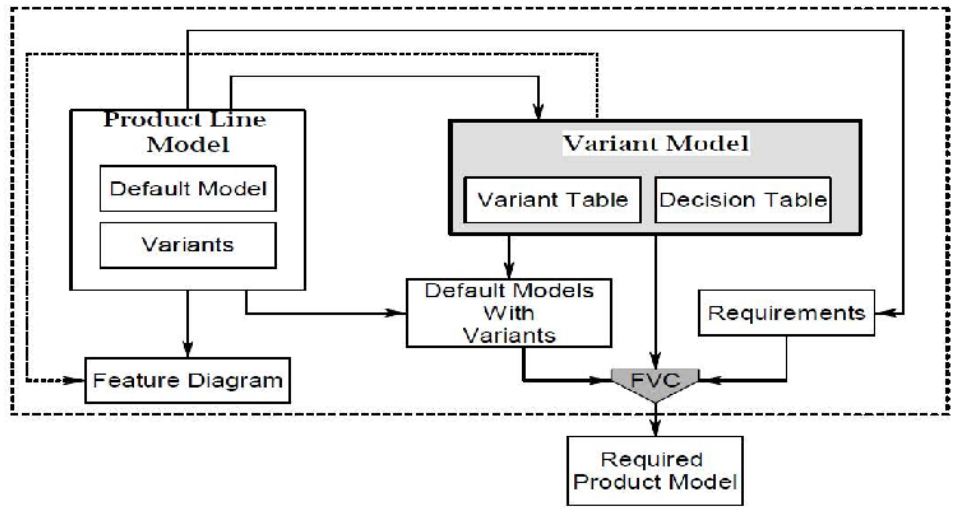

Figure 1.Variant model with product line model

The left-hand-side of Fig. 1 depicts the product line model which comprises the default model and the variants. A feature diagram can be drawn from the product line model to get an overall picture of the product line functionalities, both common and variant. The right-hand-side of the figure mainly depicts the variant model. The variant model is constructed by getting information from the product line relating to both common and variant features. A generic domain model is created by adding the variants with the default model and during itsconstruction information is collected from the variant model and also from the product line model. Finally, the required product model is developed by customizing the generic domain model after handling the variants according to the product requirements

This model carries all the variant relatedinformation like specifications, interdependencies, origins of variants, etc. UMLhas been widely used as a modelling notation for any product. However, it is onlydefined to model a single product. We use an extension mechanism of UML [5]and model the case study. In particular we use UML 'Use Case' and 'ActivityDiagram' to model the CAD domain. We then use logical representation offeature models facilitating the development of decision table in a formally soundway. We define six types of logical notation to represent all the parts in a featuremodel. First-order logic has been used for this purpose. These notations are usedto define all possible scenarios of a feature model. It is often levelled that manualverification leads to numerous error for large models and often misses the minutedetail in the verification. To overcome this problem we use the model checkerAlloy [6]. Alloy use first-order logic. We encode our logical definitions into Alloyand check the validity of the logical verification that we perform by hand. Weuse a case study of Computer Aided Dispatch (CAD) (http://xvcl.comp.nus.edu.sg/xvcl/cad/CAD.html) system product line.

In the rest of the paper, we first give a brief overview of CAD domain inSection 2. CAD variants are drawn by using FODA and UML extension andlogical notation in Section 3. Various feature analysis operations and corresponding rules are definedin Section 4. We illustrate various example configurations and describe how tologically verify them using our logical definitions. The Alloy representations oflogical verifications are outlined in Section 5. Finally, Section 6 concludes ourpaper and outlines our future plans.

## 2. CAD DOMAIN OVERVIEW

A Computer Aided Dispatch system (CAD) is a mission-critical system. Thesystem is used by police, fire and rescue, health service, port operation, taxibooking and others. Fig. 2 portrays the basic operational scenarios and roles ofa CAD system.

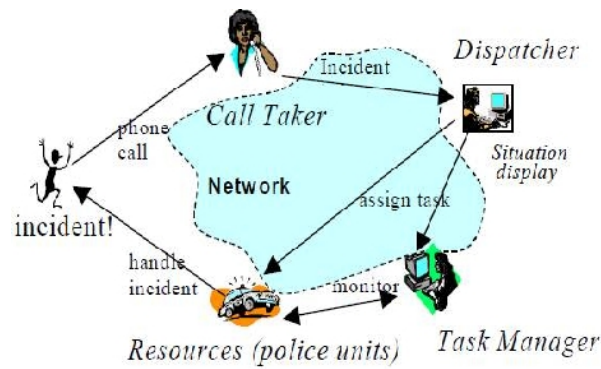

Figure 2.Basic operational scenario in a CAD system for police

After an incident, a caller reports to the command and control centre ofthe police unit. A Call Taker captures the details about the incident and theCaller, and creates a task for the incident. There is a Dispatcher in the systemwhose task is to dispatch resources to handle any incident. The system showsthe Dispatcher a list of un-dispatched tasks. After examining the situation, theDispatcher selects suitable Resources (e.g. police units) and dispatches them to execute the task. The Task Manager monitors the situation and at the end, closesthe task. Different CAD family members have different resources and tasks fortheir system.

At the basic operational level, all CAD systems are similar; basically theysupport the dispatcher units to handle the incidents. However, there are differences across the CAD systems. The specific context of operation results inmany variations on the basic operational theme. Some of the variants identifiedin CAD domain are:

- *Call taker and dispatcher roles*: In some CAD system Call taker and dispatcher roles are separated, whereas in some system their roles are mergedand one person plays the both roles.
- *Validation* of caller and task information differs across CAD systems. Insome CAD systems basic validation (i.e., checking the completeness of callerinformation and the task information) is sufficient while in other CAD systems validation includes duplicate task checking, yet in other CAD systemsno validation is required at all.
- *Un-dispatched task selection rule*: In certain situation at any given time theremight be more than one task to be dispatched and it is required to decidewhich task will be dispatched next. A number of algorithms are available forthis purpose and different CAD system use different algorithm

# 3. MODELLING SPL VARIANTS

Features are user visible aspects or characteristics of a system and are organizedinto And/Or graph in order to identify the commonalities and variants of theapplication domain. Domain features are organized into a tree-like graphical formand it is an integral part of the Feature Oriented Domain Analysis (FODA) [7]method and the Feature Oriented Domain Reuse Method (FORM) [8]. The rootnode of the tree represents the domain and the internal nodes of a tree representthe variation point and their leaves represent the values of the correspondingvariation points and known as variants. Graphical symbols are used to indicatethe categories of features such as, Mandatory, Optional, and Alternative.

Mandatory features are default part of the system. Optional features maybe selected as a part of the system if their parent feature is in the system. Alternative features, on the other hand, are related to each other as a mutuallyexclusive relationship, i.e. exactly one feature out of a set of features is to beselected. There are more relationships between features. One is Or-feature [9],which connects a set of optional features with a parent feature, either commonor variant. The meaning is that whenever the parent feature is selected then atleast one of the optional features will be selected. Feature diagram also depictsthe interdependencies among the variants which describes the selection of onevariant depends on the selection of the dependency connected variants. A CADfeature tree is illustrated in Fig. 3.

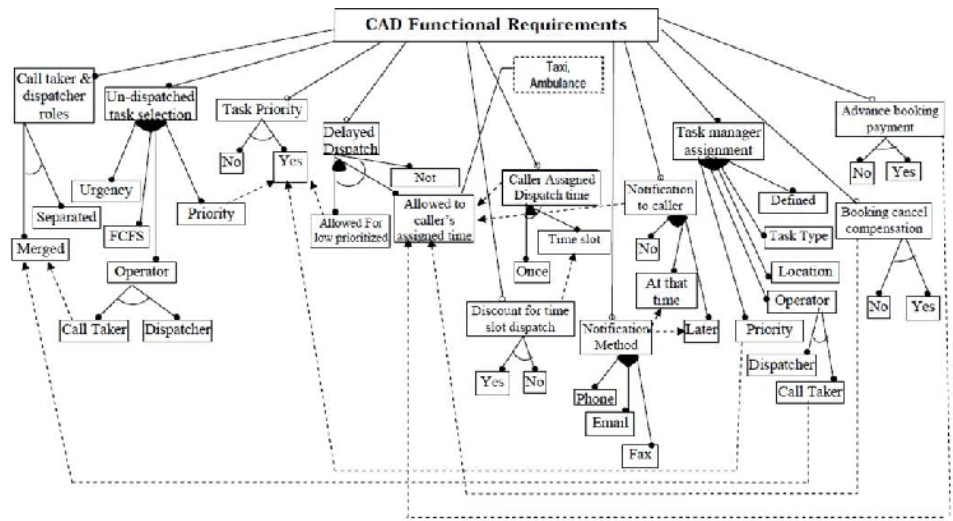

Figure 3.CAD feature diagram with dependencies

Feature models are widely used in domain analysis to model the common as well as variant requirements of the application domain. However, the semantics of a domain are not fully expressed by feature models. As a result, there is a need for other notations to support feature models which can enhance the meaning of the domain concept. The Unified Modelling Language (UML), a standardized notation for describing object-oriented models, can be used with feature model to depict the domain concept properly. UML is targeted at modelling single system rather than system families. In order to use UML diagrams to represent the model of the system family simple extension mechanisms [11] of UML, namely stereotypes and tagged values are used here. The stereotype <<variant>> designates a model element as a variant and the tagged values are used to keep trace of the models and their corresponding variant elements. It is claimed that adding only the stereotype <<variants>> does not represent the types of variants and proposed another extension where the notion of variation point is used to make variation point visible in use case diagram, represented as a triangle and variant is used to make variant in use cases explicitly.

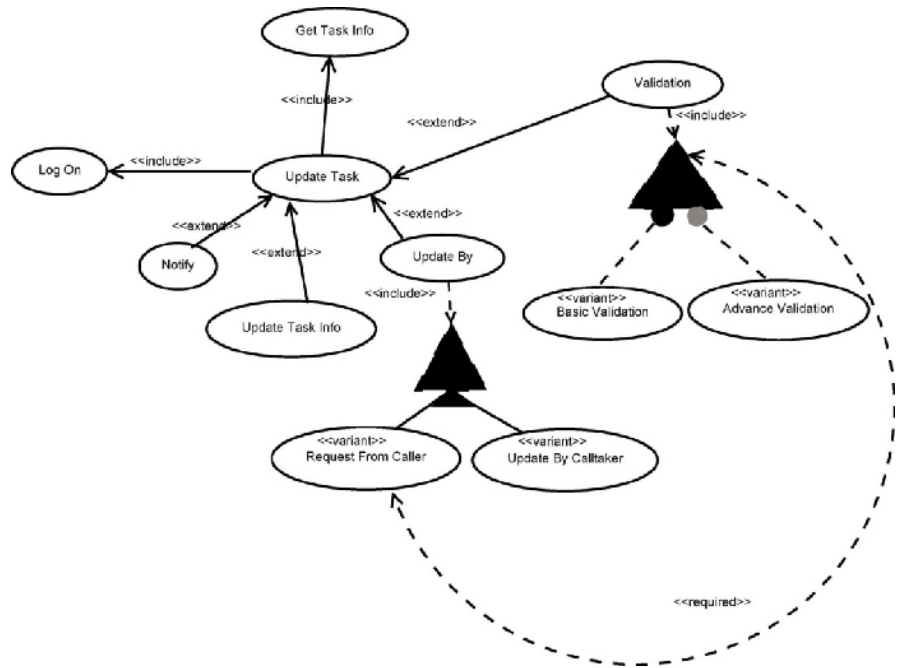

Figure 4. Update task use case

Fig. 4 illustrates the use case diagram added with variants of 'Update Task' activity. An *exclude* denotes that when one feature is selected other related feature cannot be selected. A *requires* relation indicates that when there is a relation from one feature (source) to another (target), then if the source feature is selected the target feature has to be selected as well. UML activity diagrams are used to identify the workflow of any activity. As use cases are the source of information for creating activity diagrams, whenever there is change occurs in use cases due to using <<include>> or <<extend>>, then corresponding activity diagrams should be updated. The activity diagram of a task updating a task is shown in Fig. 5.

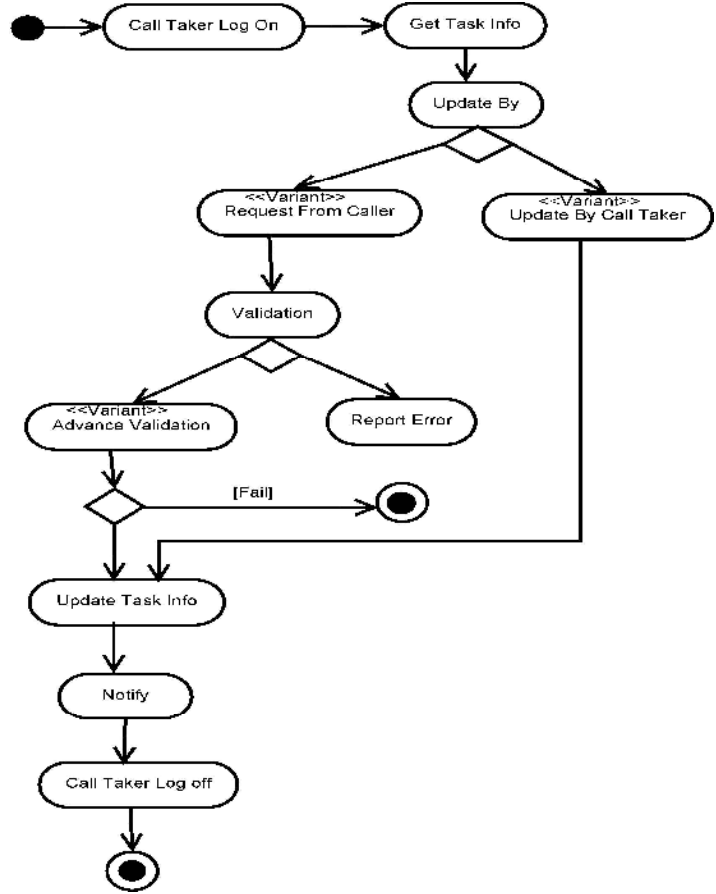

Figure 5. Update task activity diagram

A feature model is a hierarchically arranged set of features. It represents allpossible products of a software product line in a single model. The relationbetween a parent(variation point) features and its child features (variants) arecategorized as Mandatory, Optional, Alternative, Optional Alternative, Or, andOptional Or. The logical notions of these features are defined in Fig. 6.

| Type | Logic Expression | Type | Logic Expression |
|---|---|---|---|
| Mandatory | $v_p \Leftrightarrow v$ | Optional | $v \Rightarrow v_p$ |
| Alternative | $v_p \Leftrightarrow (v_1 \oplus v_2)$ | Optional Alternative | $(v_1 \oplus v_2) \Rightarrow v_p$ |
| Or | $v_p \Leftrightarrow (v_1 \vee v_2)$ | Optional Or | $(v_1 \vee v_2) \Rightarrow v_p$ |

Figure 6.Logical notations for feature models

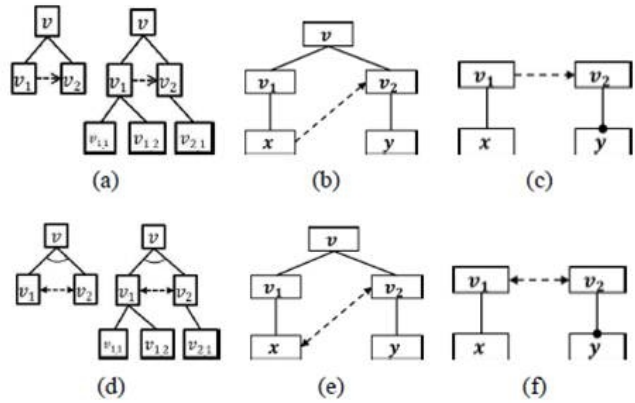

Figure 7.(a)-(c) Requires dependency between variants and (d)-(e) Exclude dependency between variants and variation points

# 4. LOGICAL ANALYSIS OF FEATURE MODEL

From the logical representation of feature model it is possible to analyse various scenarios during product customization. We consider a feature model as a graph consists of a set of subgraphs. Each subgraph is created separately by defining a relationship between the variation point (denoted as $v_i$) and the variants ($v_{i,j}$) by using the expressions shown in Fig. 5. A relationship between cross-tree (or cross hierarchy) variants (or variation points) is denoted as a dependency. There are two types of dependencies considered in this paper, inclusion and exclusion: if there is a dependency between p and q, then if p is included then q must be included (or excluded). Dependencies are drawn by dotted lines.

*Scenario 1:* If there is a `require` relation between variants $v_1$ and $v_2$ as shown in Fig. 7(a), then $v_2$ is elected whenever $v_1$ is selected. Adopting the notation in [10] the rule for dependency among variants as well as variation points is defined as follows,

$$v_1, v_2 \; type(v_1, variant) \land type(v_2, variant)$$
$$\land \; require\_v\_v(v_1, v_2) \land select(v_1) \Rightarrow select(v_2)$$
$$v_1, v_2 \cdot type(v_1, variation\_point) \land type(v_2, variation\_point)$$
$$\land \; require\_vp\_vp(v_1, v_2) \land select(v_1) \Rightarrow select(v_2)$$

Where $type(v_i)$ indicates $v_i$ is either a variant or variation point, $select(v_i)$ indicated the selection of $v_i$ and $require()$ indicates the require relationship.

*Scenario 2:* From Fig. 7(b), we derive the following rule,
$$v_1, v_2, x \cdot type(x, variant) \land type(v_1, variation\,point) \land type(v_2, variation\,point)$$
$$\land \; requires \; v_{vp(x,v_2)} \land select(x) \Rightarrow select(v_2)$$

*Scenario 3*: The following rule is derived from Fig. 7(c)
$$v_1, v_2, x, y \; \cdot \; type(x, variant) \; \land \; type(y, variant)$$
$$\land \; variants(v_1, x) \land variant(v_2, y) \land common(y)$$
$$\land \; requires \; vp_{vp(v_1, v_2)} \land select(x) \Rightarrow select(y)$$

*Scenario 4*: When there is an exclude relation between variants (and/or variation point) as shown in Fig. 7(d), only one among them can be selected at a time.

$v_1, v_2 \cdot type(v_1, variant) \wedge type(v2, variant) \wedge$
$\qquad exclude\_v\_v(v_1, v_2) \wedge select(v_1) \Rightarrow notselect(v_2)$

$v_1, v_2 \cdot type(v_1, variation\,point) \wedge type(v_2, variation\,point)$
$\qquad \wedge exclude\_vp\_vp(v_1, v_2) \wedge select(v_1) \Rightarrow notselect(v_2)$

*Scenario 5*: From Fig. 7(e) we derive the following rule,

$v_1, v_2, x, y \cdot type(x, variant) \wedge type(v_1, variation\,point)$
$\qquad type(v_2, variation\,point) \wedge exclude\,v\_vp(x, v_2)$
$\qquad \wedge select(x) \Rightarrow notselect(v_2)$

*Scenario 6*: The scenario in Fig 7(f) depicts the following rule,

$v_1, v_2, x, y \cdot type(x, variant) \wedge type(y, variant)$
$\qquad \wedge variant(v_1, x) \wedge variant(v_2, y) \quad common(y, yes)$
$\qquad requires\_vp\_vp(v_1, v_2) \wedge select(x) \Rightarrow notselect(y)$

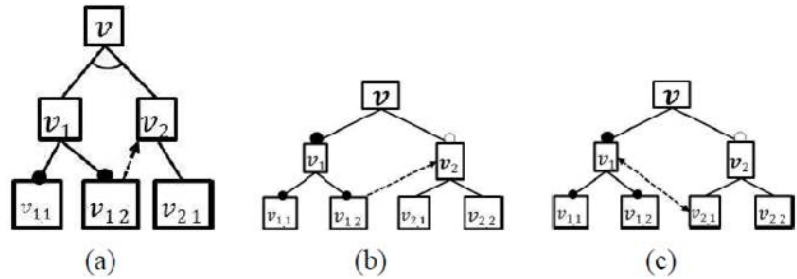

Figure 8.  (a) Inconsistency checking, (b) False optional feature detection and (c) Dead feature detection

## 4.1. Analysis Operations

We perform some analysis operations that determine whether the feature model works correctly

*Inconsistency*: In Fig. 8(a), $v, v_1$ and $v_2$ are three variation points where $v_{1.1}$ and $v_{1.2}$ are variants of $v_1$ and $v_{2.1}$ and $v_{2.2}$ are of $v_2$. There exists a require relationship between variant $v_{1.2}$ and variation point $v_2$. As $v_{1.1}$ and $v_{1.2}$ are mandatory feature whenever $v_1$ is selected both variants will be selected, and consequently, variation point $v_2$ will be selected as well due to require relation. However, $v_1$ and $v_2$ are alternative features, and both cannot be selected at the same time and it introduces an inconsistency into the feature model.

*False optional* is a situation where a feature is declared as optional which does not need to be optional. In Fig. 8(b), $v_2$ is *False optional*.

*Dead feature* is a feature that never appears in any legal product from a feature model. As shown in Fig. 8(c) due to exclude relation $v_{2.1}$ will never be part of any valid product from the feature model.

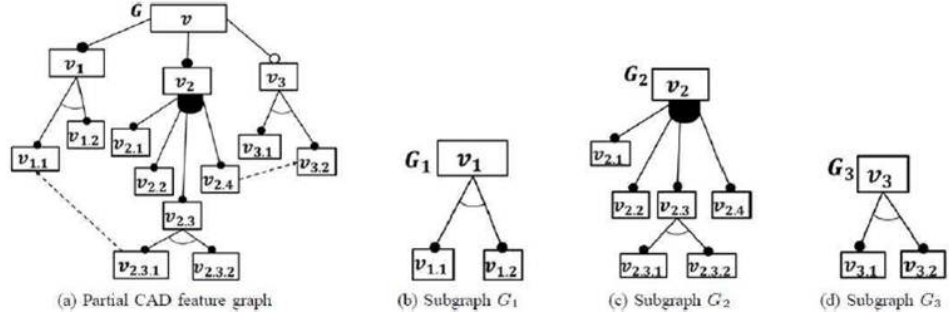

Figure 9.Partial CAD feature graph and its subgraphs

## 4.2. Analysis Examples

Automatic analysis of variants is already identified as a critical task [7]. Our logical representation can define and validate a number of analysis operations suggested in [11], [12]. In order to construct an instance product from the product line model, TRUE(T) is assigned to the selected features and FALSE(F) to those not selected. These truth values are assigned to the product line model and if TRUE value is evaluated, we call the model as valid otherwise the model is invalid. For convenience, we represent a partial tree of the CAD feature in Fig. 10 which is splitted into smaller subgraphs (Fig. 9(b), 9(c) and 9(d)).

*Example 1:* Suppose the selected variants are $v_1, v_{1.1}, v_2, v_{2.1}, v_{2.3}, v_{2.3.1}, v_{2.4}, v_3$ and $v_{3.2}$. We check the validity of the subgraphs $G_1, G_2$ and $G_3$ by substituting the truth values of the variants of the subgraphs.

$G_1 : (v_{1.1} \oplus v_{1.2}) \Leftrightarrow v_1 = T$
$G_2 : v_2 \Leftrightarrow v_{2.1} \vee v_{2.2} \vee v_{2.3} \cdot v_{2.4}$
$= v_2 \Leftrightarrow v_{2.1} \vee v_{2.2} \vee ((v_{2.3.1} \oplus v_{2.3.2}) \quad v_{2.3}) \cdot v_{2.4} = T$
$G_3 : (v_{3.1} \oplus v_{3.2}) \Leftrightarrow v_3 = T$

As the subgraphs $G_1, G_2$ and $G_3$ are evaluate to TRUE, the product model is valid. However, variant dependencies are not yet considered in this case. Checking the validity of each subgraph is not enough for the validity of the whole model. Variant dependencies must also be checked as additional constraints. We evaluate the dependencies of the selected variants and we get,

Dependency: $(v_{2.3.1} \quad v_{1.1}) \quad (v_{2.4} \quad v_{3.2})$
$= (T \Rightarrow T) \quad (T \quad T) = T$

The truth (T) value of the dependencies ensures the validity of the product instance

*Example 2*: Suppose the selected variants are, $v_1, v_{1.2}, v_2, v_{2.3}, v_{2.3.1}, v_{2.4}, v_3, v_{3.1}$. To check whether these input combination build a valid product we check the validity of the sub-graph $G_1, G_2$ and $G_3$ by substituting the truth values of the variants of the sub-graphs.

$G_1 \quad (v_{1.1} \oplus v_{1.2}) \Leftrightarrow v_1 = T$
$G_2 \quad (v_{2.1} \vee v_{2.2} \vee v_{2.3} \vee v_{2.4}) \Leftrightarrow v2$
$= (v_{2.1} \vee v_{2.2} \vee ((v_{2.3.1} \oplus v_{2.3.2}) \quad v_{2.3}) \cdot v_{2.4} \quad v_2 = T$
$G_3 \quad (v_{3.1} \oplus v_{3.2}) \Leftrightarrow v3 = T$

We then evaluate the dependencies of the selected variants,

*Dependency*: $(v_{2.3.1} \quad v_{1.1}) \quad (v_{2.4} \quad v_{3.1}) = F$

Due to conflict within variant dependencies, the whole graph becomes invalid, which is due to an incorrect selection of input.

## 5. ALLOY VERIFICATION

In order to define feature model in Alloy we first define the abstract syntax ofthe feature model. A feature model (FM) has a set of features and one rootfeature. A FM also has a set of relations and formulas.

```
sig FM {
      features: set Name,
      root: feature,
      relation: set Relation,
      form: Formula
}
abstract sig Type{}
one sig Optional, Mandatory, OrFeature,
Alternative OptionalAlternative, OptionalOr
extends Type {}

sig Name{}
sig Relation{
parent: Name,
child: set Name,
type: Type
}
```

Feature model declares formulas using propositional logic that returns aBoolean value when a configuration satisfies a formula. An Alloy signature isalso declared for binary formulas.

```
abstract sig Formula
{
satisfy: configuration
->one Bool
}
abstract sig Op{}
one sig AndF, OrF, Implies, NotF
extends Op{}
Sig Form extends Formula {
f: Formula,
g: Formula,
op: Op
}
```

After defining the abstract syntax, the semantics are defined by first defining the configuration which is a set of feature names. The semantics of FM isthen defined as the set of configurations that satisfy all the relations whereasconstraints are denoted as predicates. We also define several constraints overconfigurations as well as rules for formula satisfactions.

*Example*: We show how the feature diagram in Fig. 9(a) can be constructedfrom the syntax and semantics defined in previous section. The feature diagramconsists of three subgraphs$G_1, G_2$and $G_3$. First we define the overall diagramby using the subgraphs and their relations and then define the parent-childrelations.

```
one sig CAD extends FM{}
one sig c1,c2,c3 extends Relation{}
fact elements { CAD.root = v
CAD.feature = G1 + G2 + G3
CAD.relation = c1 + c2 + c3 }

fact relations{
c1.type = Mandatory
c1.parent = v c1.child = v1
c2.type = Mandatory
c2.parent = v c2.child = v2
c3.type = Optional
c3.parent = v c3.child = v3 }
```

Alloy checks the consistency of the sub-graphs and variant dependencies,and displays that a valid instance is found which is indicated bythe alloy result display screen, where it shows that an instance is found as shown in Fig.10.

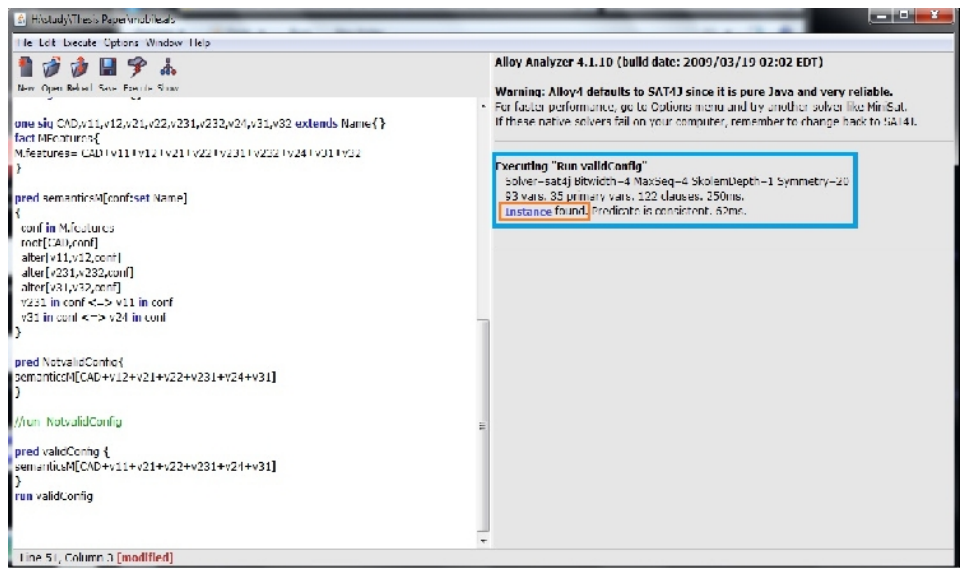

Figure 10.Valid instance check in Alloy

If we select another combination of features as in Example 2 in earlier section, an invalid product would be created and hence in instance should be found. Alloy produce an error stating that no instance is found from the selected features as shown in Fig 11.

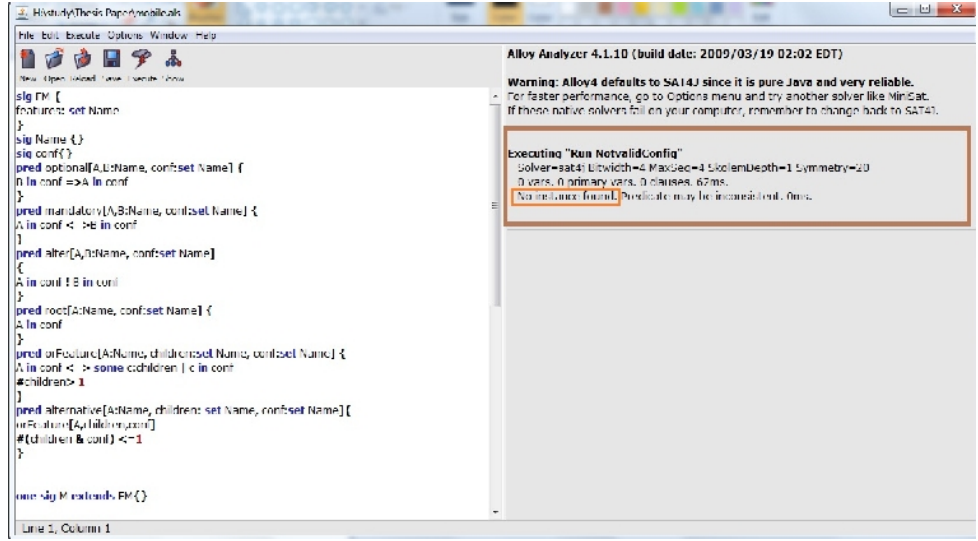

Figure 11.Invalid instance checking in Alloy

Currently, we are defining explicitformulas using our Alloy definitions. Our plan is first to define the whole featurediagram then define the formulas specifying the relations between features andoperators.

# 6. CONCLUSIONS

Modelling variants is considered as one the crucial factor for the successful deployment of software product line. In our systematic modelling, first the variantsare visually arranged in a feature diagram that illustrates various relationshipsamong the features. Although UML is designed for single systems, an extendedversion of UML has been used in this paper to model SPL variants. To be able toformally verify the variant configuration and consistency we model all six typesof variant relations by using first-order logic. Cross tree variants dependencies aredefined as well. Such formal definition facilitates the automated decision makingduring product customization. The various analysis operation suggested in [11,12] are also addressed here. To overcome the hurdles of by-hand verification, weencode our first-order logic representation into Alloy [6], that automaticallychecks the consistency of feature configuration and validity of product instances.